\documentclass[useAMS,usenatbib]{mn2e}

\usepackage{graphicx}

\title[Dark Matter Cusps in Clusters]{Shallow Dark Matter Cusps in Galaxy Clusters}

\author[Laporte et al.]{
\parbox[t]{\textwidth}{
Chervin F. P. Laporte$^{1}$,
Simon D. M. White,$^{1}$
Thorsten Naab,$^{1}$
Mateusz Ruszkowski$^{2,3}$ and
Volker Springel$^{4,5}$
}
\\
\\
$^{1}$ Max Planck Institute for Astrophysics, Karl-Schwarzschild-Strasse 
1, 85740 Garching, Germany\\
$^{2}$ Department of Astronomy, University of Michigan, 500 Church St., Ann Arbor MI 48109, USA\\
$^{3}$The Michigan Center for Theoretical Physics, 3444 Randall Lab, 450 Church St, Ann
Arbor, MI 48109, USA\\
$^4$Heidelberger Institut f\"{u}r Theoretische Studien, Schloss-Wolfsbrunnenweg 35, 69118 Heidelberg, Germany\\
$^5$Zentrum f\"ur Astronomie der Universit\"at Heidelberg, Astronomisches Recheninstitut, M\"{o}nchhofstr. 12-14, 69120 Heidelberg, Germany\\
}

\begin{document}
\date{}
\pagerange{\pageref{firstpage}--\pageref{lastpage}} \pubyear{2011}
\maketitle
\label{firstpage}
\begin{abstract}
We study the evolution of the stellar and dark matter components in a galaxy cluster of $10^{15} \, \rm{M_{\odot}}$ from $z=3$ to the present epoch using the high-resolution collisionless simulations of Ruszkowski \& Springel (2009). At $z=3$ the dominant progenitor halos were populated with spherical model galaxies with and without accounting for adiabatic contraction. We apply a weighting scheme which allows us to change the relative amount of dark and stellar material assigned to each simulation particle in order to produce luminous properties which agree better with abundance matching arguments and observed bulge sizes at $z=3$. This permits the study of the effect of initial compactness on the evolution of the mass-size relation. We find that for more compact initial stellar distributions the size of the final Brightest Cluster Galaxy grows with mass according to $r\propto M^{2}$, whereas for more extended initial distributions, $r\propto M$. Our results show that collisionless mergers in a cosmological context can reduce the strength of inner dark matter cusps with changes in logarithmic slope of 0.3 to 0.5 at fixed radius. Shallow cusps such as those found recently in several strong lensing clusters thus do not necessarily conflict with CDM, but may rather reflect on the initial structure of the progenitor galaxies, which was shaped at high redshift by their formation process.
\end{abstract}
\begin{keywords}
galaxies: formation - galaxies: evolution - galaxies: clusters: general - galaxies: elliptical and lenticular, cD
\end{keywords}

\section{Introduction}

Brightest cluster galaxies (BCGs) are the most massive elliptical galaxies in the Universe at the extreme end of the galaxy luminosity function and perhaps ``special'' (see \citealp{Sheth2011}). These objects are intriguing because they do not follow the same scaling relations as normal giant ellipticals.  Using the Sloan Digital Sky Survey (SDSS), \cite{vonderLinden2007} found that BCGs deviate systematically from the \cite{Faber1976} and the \cite{Kormendy1977} relations with lower velocity dispersions and larger sizes respectively. Early theoretical studies investigated the role of cooling flows \citep{Fabian1994} in BCG formation. This hypothesis is disfavoured by observations with Chandra and XMM-Newton which show that such flows are much weaker than required, as are the star formation rates in the central galaxies of clusters \citep{McNamara2000, Fabian2001}. Another scenario has BCGs growing by feeding on smaller galaxies (minor mergers). This is the notion of galactic cannibalism \citep{Ostriker1977, Hausman1978}. 

In the $\Lambda$CDM scenario, structures build-up hierarchically through accretion and mergers of smaller progenitors. Groups form before clusters and have sufficiently low relative velocities that galaxy-galaxy mergers can occur before cluster formation, thus enhancing the formation of a massive central galaxy. \cite{DeLucia2007} use semi-analytic galaxy formation models to show that in the $\Lambda$CDM cosmology, BCGs form primarily through in-situ star-formation at high redshifts, $z\geq3$, with subsequent mass growth dominated by non-dissipational merging. Similar processes are seen in hydrodynamical cosmological simulations of massive galaxy formation \citep{Naab2009,Oser2011, Feldmann2011} although these are typically less effective at suppressing star formation at late times than semi-analytic models, leading to galaxies which are ``younger'' than those observed.

From this, it seems that many aspects of the late assembly of BCGs can be modelled without considering the early star formation phase. \cite{Dubinski1998} was an early example of such work based on numerical N-body simulations of collisionless mergers of galaxies in a cosmological context. Dubinski found that the properties of cluster BCGs can be naturally explained by merging of galaxies which have already formed their stars at high redshift. This idea was further investigated by \cite{Ruszkowski2009} (RS09) who studied the deviations of BCGs from the Kormendy and FJ relations in a $\Lambda$CDM simulation.

In relaxed clusters, BCGs reside at the bottom of the potential well, making them ideal probes of the distribution of dark matter from kpc to Mpc scales. \cite{Sand2002,Sand2004, Sand2008} studied a selection of clusters combining stellar dynamical modelling and strong gravitational lensing in order to infer the inner slope of the dark matter density profiles. Their studies found values for the logarithmic slope of the dark matter density profile $\gamma=-d\ln(\rho)/d\ln r < 1$, at odds with the predictions of dark-matter-only simulations of halo formation which generally follow the NFW profile with $\gamma_{\rm NFW}= 1$ \citep{Navarro1997}. More recently, \cite{Newman2011} revisited the study of Abell 383 by \cite{Sand2008}, combining stellar kinematics, strong and weak lensing and X-ray data to deduce an inner slope of $\gamma=0.59^{+0.30}_{-0.35}$ at 95 percent confidence. These authors suggest this may indicate a genuine problem with our understanding either of baryonic evolution or of the nature of the dark matter.

Indeed, some recent hydrodynamical simulations of galaxy formation in clusters indicate a steepening of the inner-slope of the dark matter profile \citep{Gnedin2004, Gnedin2011, Sommer-Larsen2010}, exacerbating the core-cusp problem . More gentle contraction is seen in other simulations (e.g. \citealp{Duffy2010}) but not on the scales investigated by \cite{Newman2011}. A simulation in which dynamical friction significantly erodes the cusp at the center of ellipticals was presented by \cite{Johansson2009} but on the scale of the central galaxy of a small group. However, the final masses of the central galaxies in these simulations are in general a factor of two or three higher than expected from abundance matching arguments \citep{Guo2010, Moster2010, Behroozi2010}, implying the need for a signicantly improved treatment of baryonic astrophysics. \cite{El-Zant2004} claimed that shallower cusps could be produced in a cluster through dynamical heating by the galaxies. However, they treated galaxies as unstrippable point masses which is too unrealistic to address the issue in quantitative detail.

At this stage it still seems interesting to address the second question by \cite{Newman2011}: is the presence of shallow dark matter cusps at the centre of clusters a significant challenge to CDM? We use the RS09 simulation to test whether such cusps can be created through dry (i.e., gas-free) mergers. Recent observations of massive ellipticals at $z=2$ have shown that they were more compact than similar mass galaxies today (see e.g. \citealp{Dokkum2008}). Dry, predominantly minor mergers have also been proposed as a possible mechanism to drive the required size evolution (e.g., \citealp{Naab2009, Bezanson2009}) 

 In this context a significant limitation of the RS09 simulations was that the galaxies they inserted at $z=3$ had stellar masses an order of magnitude larger than expected from abundance matching arguments \citep{Moster2010} and were assumed to follow the present-day mass-size relation. Here we remedy the inconsistencies between the simulations and observations by using a method that re-assigns the mixture of stellar and dark matter in each simulation particle. This enables us to study the evolution of stellar and dark matter distributions for different levels of initial compactness and stellar mass.

In Section 2, we give a description of the simulations as well as of the weighting scheme used in this study. We also present results on the mass and size growth of the BCG for different initial assignments of stars and dark matter. In Section 3, we look at how the initial slope of the dark matter evolves from $z=3$ to the present. We discuss our results and conclude in Section 4.

\section{Numerical Methods}
\subsection{Simulation}
The simulations used for this study are described in detail in \cite{Ruszkowski2009}, (RS09) and we give only a short
summary here. A cluster mass dark matter halo of $10^{15} \, {\rm{M_{\odot}}}$ was identified in the 
{\it Millennium Simulation} \citep{Springel2005b}. The cosmological parameters of this simulation are
 $\Omega_{m} = 0.25$, $\Omega_{\Lambda} = 0.75$, a scale-invariant slope of the power spectrum of 
primordial fluctuations ($n = 1.0$), a fluctuation normalization $\sigma_{8} = 0.9$, and a Hubble constant
$H_{0} = 100 \, h \, {\rm{km\,s^{-1} \, Mpc^{-1}}} = 73 \, {\rm{km\,s^{-1} \, Mpc^{-1}}}$.

The cluster was then re-simulated using a zooming technique with a mass resolution of $m=1.57\times 10^{7} h^{-1}\,
{\rm M_{\odot}}$ and comoving softening length $\epsilon=2.0 \, h^{-1} \, {\rm kpc}$. At redshift $z=3$, the 50 most
massive progenitors of the final cluster were identified and replaced by spherical equilibrium models in which stars were distributed according to a  \cite{Hernquist1990} profile embedded in a NFW dark matter
halo \citep{Navarro1997}. The masses of the dark matter and star particles were set to be identical 
$m_{dm}=m_{*}=1.57 \times \, 10^{7} \, h^{-1} \,  {\rm{M_{\odot}}}$ and they were assigned a softening length of $\epsilon=1 \, h^{-1} \, {\rm kpc}$.
 comoving, half that of the other dark matter particles from the original resimulation. Throughout this paper we use $h=0.73$ and our mass and length units are thus in $\rm{kpc}$ and $\rm{M_{\odot}}$.

Two simulations were run with different initial galaxy models: one in which the dark matter was adiabatically contracted following \cite{Blumenthal1986} and one where it retained an undisturbed NFW profile. We shall refer to these as models A and B respectively. The \cite{Blumenthal1986} formalism over-predicts the amount of contraction observed in many hydrodynamical simulations \citep{Gnedin2004}, however through the inclusion of uncompressed and compressed 
dark halo models, we can probe two alternative regimes. If dry-merging is indeed the main driver in the late 
assembly of BCGs and if the RS09 initial galaxies were realistic, then the ``real" solution would lie between these two models. 

In fact, however, the galaxies which RS09 inserted at redshift $z=3$ followed the {\it present-day} mass-size relation from \cite{Shen2003} and assumed a stellar to dark matter mass ratio $m_{*}/M=0.1$. This value is too large by a factor of 10 according to recent results from matching the observed high-redshift abundance of massive galaxies \citep{Moster2010, Behroozi2010, Wake2011}. We note that many other simulations investigating similar processes have like-wise assumed over-massive stellar components (e.g., \citealp{Nipoti2009,Rudick2011}). However, we point out that for a given mass resolution, more massive stellar bulges are represented by a larger number of particles which considerably improves the numerical convergence of the simulations.

 Not surprisingly, the final merger remnants in RS09 were also too massive $m_{*}\sim10^{13} \, \rm{M_{\odot}}$ with half-mass radii which were too large ($\sim100 \, \rm{kpc}$) compared to real BCGs. These generally do not exceed $r_{e}\sim 50 \rm{kpc}$ \citep{Bernardi2007}. 

In order to study more consistently the change in the slope of dark matter density profiles at the centre of $\Lambda$CDM clusters, we need to address the question with galaxies that have stellar properties consistent with observations. We also need to test whether the final stellar mass of the merger remnant agrees with the stellar to halo mass relation (SHM) at $z=0$ \citep{Guo2010, Moster2010, Behroozi2010}. We employ a weighting procedure to re-assign the luminous component of every initial galaxy such that only one percent of its total dark matter mass is locked in stars. For the range of halo masses ($10^{13} \, \rm{M_{\odot}} - 10^{12} \, \rm{M_{\odot}}$) that we populate, abundance matching implies that this ratio is rather constant. Additionally, our weighting scheme enables us to change the sizes of the luminous components to study the assembly of the BCG for different levels of compactness while keeping the total initial stellar mass fixed. We present this scheme in the next section.

\subsection{Weighting Scheme}

\begin{figure}
\includegraphics[width=0.5\textwidth,trim=0mm 0mm 0mm 0mm,clip]{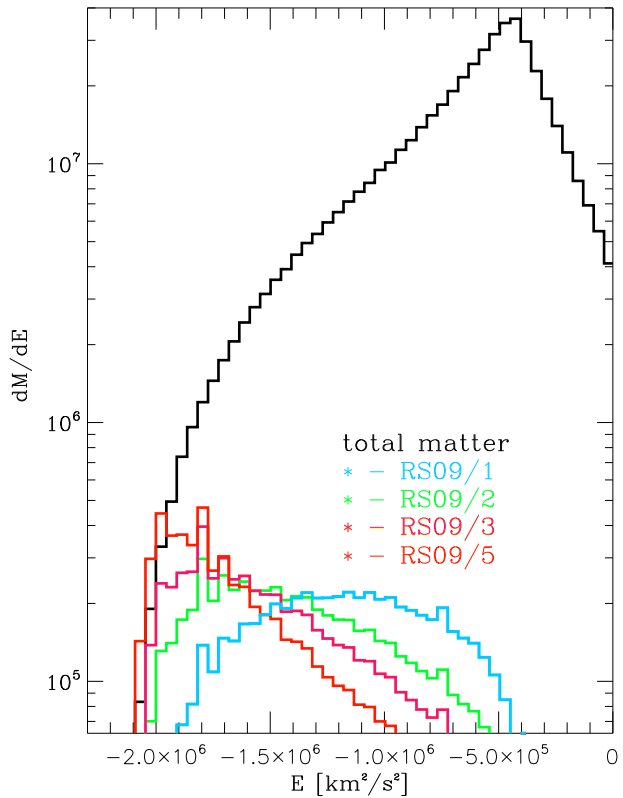}
\caption{Differential energy distribution for the proto-BCG for the total mass and light contributions for three test cases $r_{e}=r_{e_{\rm RS09}}/1,r_{e_{\rm RS09}}/2,r_{e_{\rm RS09}}/3, r_{e_{\rm RS09}}/5$. Although the $r_{e_{\rm RS09}}/5$ histogram intersects the total differential energy distribution, the particles in those energy bins are below our spatial resolution of $\epsilon=1\, h^{-1} \,\rm{kpc}$.}
\end{figure}

\begin{figure}
\includegraphics[width=0.5\textwidth,trim=0mm 0mm 0mm 0mm,clip]{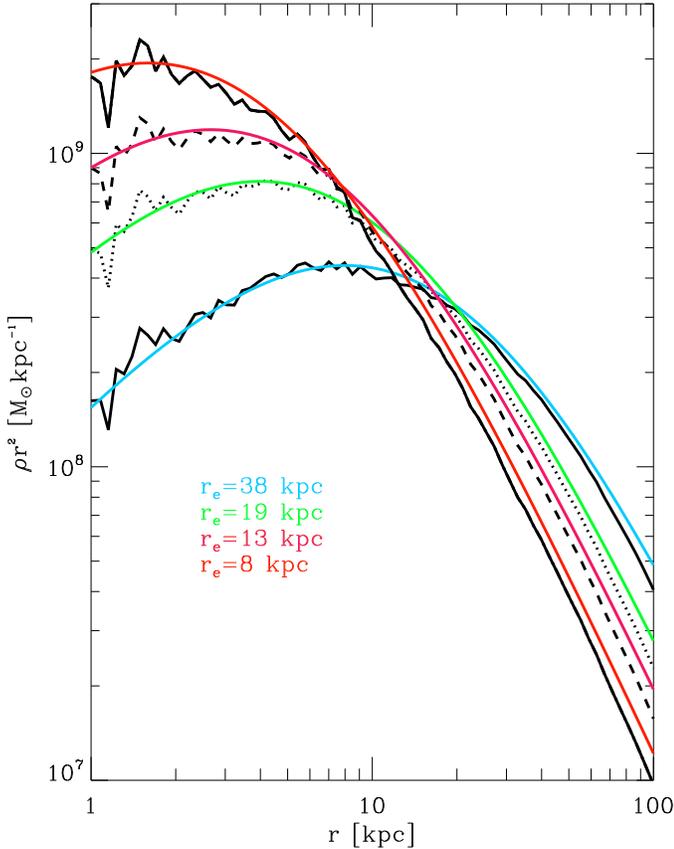}
\caption{$\rho r^2$ profiles vs. radius for the four different realisations. The target functional forms of each realisation are overplotted in solid colored lines.}
\end{figure}
Here we describe our method for re-assigning simulation particles in the initial conditions so that the light to stellar mass ratio and the size of the luminous component of a galaxy can be varied. Our scheme is similar to that of \cite{Bullock2005} and is applicable to spherically symmetric density distributions. A galaxy is represented by a distribution of $N$ particles of mass $M_{\rm tot}/N=m_{p}$ and phase-space coordinates $(\mathbf{x},\mathbf{v})$ which generate the potential $\Phi=\Phi_{\rm dark}+\Phi_{\rm stellar}=\Phi_{\rm NFW}+\Phi_{\rm Hernquist}$. Each particle of energy $E=\frac{1}{2}v^{2}+\Phi$ now simultaneously represents dark matter and stars in different amounts according to a weight function $\omega(E)=f_{*}(E)/f(E)$. This is the ratio between the stellar distribution function ({\sc df}) and the total {\sc df}. One can construct the stellar {\sc df} in the following way.

In computing $f_{*}(E)$, one assigns the particles a spherical number density distribution $\nu(r)$ and solves the Eddington inversion formula:
\begin{equation}
f_{*}(\mathcal{E})=\frac{1}{\sqrt{8}\pi^2} \int^{\mathcal{E}}_{0}\frac{d\Psi}{\sqrt{\mathcal{E}-\Psi}}\frac{d^2\nu(\Psi)}{d\Psi^2} + \frac{1}{\sqrt{\mathcal{E}}}\frac{d\nu}{d\Psi}\bigg|_{\Psi=0},
\end{equation}
 where  $\Psi=-\Phi+\Phi_{0}$ and $\mathcal{E}=-E+\Phi_{0}=\Psi-v^2/2$ are the relative potential and total energies respectively. The potential used to generate the model galaxies in RS09 was a linear combination of a Hernquist and an NFW potential, both of which tend to zero in the limit $r$ goes to infinity thus $\Phi_{0}=0$.
We choose $\nu(r)$ to follow a Hernquist profile (1990):
\begin{equation}
\nu(r)=\frac{a}{r(r+a)^3}.
\end{equation}

The total {\sc df} is the ratio of the differential energy distribution $N(E)=dM/dE=f(E)g(E)$ and the density of states $g(E)$ which is solely defined by the potential $\Phi$:
\begin{equation}
g(E)=(4\pi)^2\int^{r_{E}}_{0}r^{2}\sqrt{2(E-\Phi(r))}dr.\\
\end{equation}
Since $\Phi(r)$ and $N(E)$ can both be measured directly from the simulation's initial condition, this determines $f(E)$.

Note that the only free parameter we have introduced is the scale radius $a$ for the light distribution which is related to the half-light radius by $a=r_{e}/(\sqrt{2}+1)$. Within certain limits we can vary the relative mixture of dark and luminous matter to represent, for example, less massive and more compact bulges. This has the advantage of allowing us to study aspects of the assembly of massive galaxies from $z=3$ to $z=0$ without having to run additional CPU intensive simulations, simply by tracking the weights to the final merger remnant.
This scheme allows multiple interpretations of a single simulation. However, for our purposes the stellar mass within each galaxy is kept fixed, its value being dictated by abundance matching arguments, so it is the stellar and dark matter distributions which vary with the total mass distribution held fixed. Note that this implies that the initial dark matter distributions in the adjusted galaxies are no longer those expected naturally in $\Lambda$CDM. Thus while we can address issues of how the mixing of the two components changes inner profile shapes and is affected by initial compactness, we cannot expect the final DM distributions to be realistic.

Figure 1 shows the differential energy distribution for the most massive of the $z=3$ galaxies (the dark halo and stars of the main BCG progenitor) as a function of energy. Overplotted are stellar differential energy distributions for four reinterpretations with different galaxy sizes: the original effective radius used in RS09 and the same reduced by factors of two, three and five. To show more explicitly that our method works we present $\rho r^{2}$ profiles in Figure 2. Note that changing the sizes of the stellar component for a fixed $m_{*}/M_{\rm halo}$ ratio, implies substantial changes in the inner slope of the dark matter profile. The reduction of the stellar mass by a factor of 10 from that assumed in RS09 also means that both the uncontracted and contracted models now have overly concentrated dark matter distributions in the centres of the galaxy subhalos, except at very small radii where the reduced radii can lead to an increase of stellar density relative to RS09.

The maximum extent to which we are able to rescale the stellar component is set by the total mass profile. This is saturated by the stars alone at the softening radius if the RS09 sizes are reduced by a factor of $\sim5$ (see Figure 1).

The initial and final light profile shapes in the RS09/1 interpretation will be the same as in the original simulation, as only the stellar masses of every galaxy are changed (the inner dark matter profiles will differ, however, since they now contain the additional mass which used to be assigned to stars). This also means that the galaxies no longer lie on the \cite{Shen2003} stellar mass-size relation. In order to put them back on it (within the scatter) we need to reduce the sizes by factors of $\sim5$.

Recent observations show, however, that $z=2$ elliptical galaxies were more compact than implied by the local relation \citep{Dokkum2008}. Unfortunately, our spatial resolution limit does not permit us to consider such small sizes. We stress that these observations still need to be treated with caution as the galaxy stellar masses are estimated photometrically. \cite{Martinez2011} argue that dynamical masses of compact galaxies at redshift $z=1$ may be six times lower than some photometric estimates. Nevertheless if the photometrically determined stellar masses of galaxies at redshift $z=2$ are even approximately correct, the galaxies should be even smaller than we assume in this paper. As we will see, the exercise presented here can nonetheless give insight into the puffing-up of  BCGs by minor mergers and its dependance on the initial compactness of the galaxies.

\subsection{Results for the BCG evolution}
\subsubsection{Size growth of the BCG}

Fixing the stellar masses within all haloes according to abundance matching arguments (and hence reducing them by an order of magnitude from those originally assumed by RS09), we studied four assumptions for the compactness of the galaxies $r_{e}=r_{e_{\rm RS09}}/1,r_{e_{\rm RS09}}/2,r_{e_{\rm RS09}}/3, r_{e_{\rm RS09}}/5 $  in each of our two simulations. The trends arevery similar in the two models. We find that the relative growth in size of the BCG from $z=3$ to $z=0$ is not identical for different assumptions about initial concentration. This is illustrated in Table 1, where we give the growth factors as characterised by the increase in half-light radius. The size of the BCG at $z=0$ in the original RS09 simulations was high by a factor of two compared to local BCGs. However, in the RS09/5 case, the sizes of the initial galaxies correspond to those of low-redshift largest ellipticals of similar stellar mass (within the scatter of the Shen et al. 2003 relation) and the most massive BCG progenitor grows by a factor of 4, reaching a size that is in reasonable agreement with observed BCGs. This supports the idea that massive ellipticals can grow rapidly in size through repeated minor mergers in a cosmological context \citep{Naab2009, Oser2010, Oser2011, Shankar2011}.

\subsubsection{Mass growth of the BCG}
Turning to the mass growth of the BCG, the main progenitor starts with a stellar mass of $\sim2.5 \times \, 10^{11} \rm{M_{\odot}}$ at $z=3$ and the final merger remnant at $z=0$ has a total stellar mass of $6$ to $7 \times \,10^{11}\rm{M_{\odot}}$. For the cluster mass we are considering here ($\sim10^{15}\, \rm{M_{\odot}}$), abundance matching suggests $m_{*}/M_{\rm halo}=0.001$ (see Fig. 2 of \cite{Guo2010}). Considering that semi-analytic models predict that 80 percent of the stars which end up in a BCG are already formed by $z=3$ \citep{DeLucia2007} our value of $m_{*}/M(z=0)=0.0006$ is a considerable improvement over the value prior to rescaling ($m_{*}/M(z=0)\sim0.01$). { We note that this factor is still somewhat lower than expected from semi-analytic models. A quick query in the Millenium database indicates that only 30 percent of the stars in the $z=0$ BCG in the cluster come from the 50 most massive progenitor haloes at $z=3$. The missing contribution comes from smaller subhaloes which were not populated with stars in the simulation. Figure 3 shows the normalised radial distribution of simulation BCG stars coming from different galaxies at $z=0$. The two panels illustrate the process of mass aggregation for two extreme interpretations (RS09/1 and RS09/5). The stars coming from the proto-BCG (in-situ) are represented by the solid black line and those accreted from other galaxies by the dashed-dotted line. Contributions from individual galaxies are shown by coloured lines. Between $z=3$ and $z=0$ the BCG was subjected to two mergers with mass ratios of about 3:1 and six with about 10:1. These are responsible for the inside out growth of the BCG.

For the most compact galaxy interpretation (RS09/5), 30 percent of the BCG mass comes from accretion which dominates beyond $\sim 30 \rm{kpc}$. This is in contrast with the extended galaxy (RS09/1) where this transition occurs at $60 \rm{kpc}$. Note that little of the accreted mass reaches the centre. Most gets deposited on the outskirts leading to inside-out growth.

The reason why the mass growth factor decreases with increasing compactness is that the initial compact galaxies have more stars on higher binding energy orbits than
their less compact counterparts (as can be seen in Figure 1). These stars are harder to unbind. Thus, as encounters take place, less stellar mass is stripped and deposited on the BCG. More remains bound to the original galaxies.

Another interesting aspect of our numerical experiment is the scaling between total stellar mass and effective radius. For the most extended galaxies, the BCG radius and mass scale as $r_{e}\propto M$. For the compact galaxies in the RS09/5 interpretation this scaling is almost $r_{e}\propto M^2$. This latter scaling agrees with  \cite{Dokkum2010} who studied the growth of compact galaxies at fixed number density finding $r_{e}\propto M^{2.04}$. Thus for smaller sizes, we get a stronger size evolution in better agreement with observation. Seeing that our weighting scheme produces central galaxies with appropriate stellar masses for their host dark halo, we now address the issue of the change in inner slope of the dark matter profile in the presence of a baryonic component, and how this varies depending on the relative distribution of stars and dark matter.

\begin{table}
 \centering
 \begin{minipage}{130mm}
  \begin{tabular}{@{}llrrrrlrlr@{}}
  \hline
Run & $r_{e}/\rm kpc$ & $r_{e_{f}}/r_{e_{i}}$A & $M_{f}/M_{i}$A & $r_{e_{f}}/r_{e_{i}}$B  & $M_{f}/M_{i}$B\\
  \hline
RS09/1   & 38  & 2.7 & 2.9 & 2.5 & 2.9\\
RS09/2 & 19  & 3.2 & 2.7 & 3.0 & 2.6 \\
RS09/3 & 13  & 3.6 & 2.5 & 3.1 & 2.5 \\
RS09/5 & 8 & 4.3 & 2.4 & 3.8 & 2.4\\
\hline
\end{tabular}
\end{minipage}
\caption{Initial half-light radii $r_{e_{i}}$, size and mass growth factors (defined as the ratio of initial and final half-light radii and stellar masses) for the eight interpretations. The initial stellar mass of the largest BCG progenitor is $2.5 \times \, 10^{11} \rm{M_{\odot}}$ in all cases.}
\end{table}

\begin{figure}
\includegraphics[width=0.5\textwidth,trim=0mm 0mm 0mm 0mm,clip]{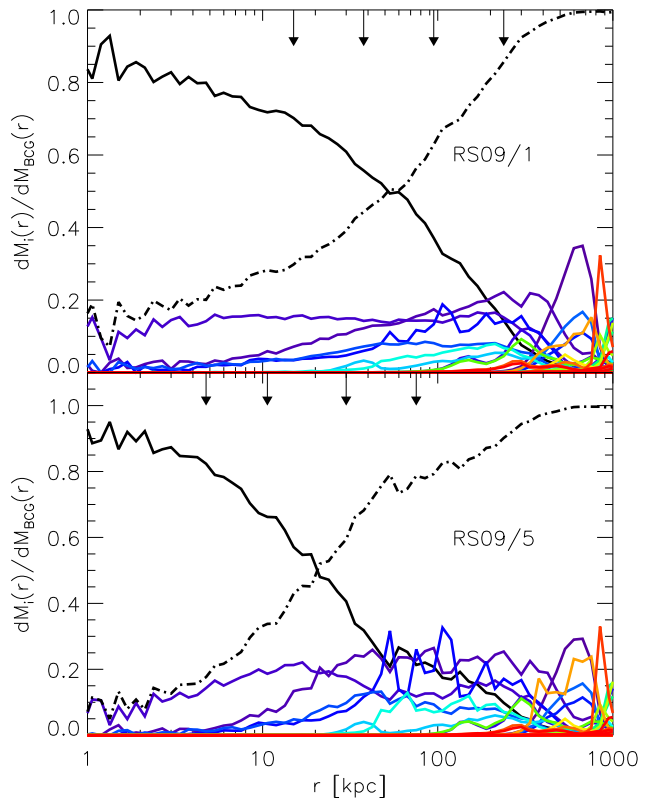}
\caption{Normalised radial distribution of stellar mass in the $z=0$ BCG coming from different contributing galaxies for model A (each colour). The solid line shows the stars coming from the largest progenitor, taken to be the proto-BCG (in-situ). The dashed-dotted line is the radial distribution of stellar mass contributed by accreted galaxies. The top and bottom panels show the results for two interpretations (RS09/1 and RS09/5). In the RS09/5 case, accreted material contributes 50 percent of the mass budget already at 30 kpc while this occurs at $60 \rm \, kpc$ for the RS09/1 case. The arrows point out the radii containing 10, 25, 50 and 75 percent of the BCG stars.}
\end{figure}

\section{Evolution of the dark matter slope}
\subsection{Methodology}

In relaxed clusters, BCGs are usually coincident with the centre of the cluster, defined theoretically as the bottom of its gravitational potential well and empirically as the centre of the X-ray emitting hot gas. Thus we define the centre of the galaxy (BCG) and cluster as the position of the particle with the minimum potential identified by the {\sc Subfind} algorithm \citep{Springel2001}.We then compute  density profiles in spherical shells around this centre, using 44 bins spaced logarithmically between $0.1 \rm{kpc}$ and $2500 \rm{kpc}$. The intrinsic slopes of the density profiles are computed by numerical differentiation using a 3-point Lagrangian interpolation as in \cite{Navarro2010}.

\subsection{Results for the original RS09 simulations}

We begin by analysing the simulations as originally presented in RS09. In Figure 4, we present the intrinsic logarithmic slope $\gamma=-d\ln(\rho)/d\ln r$ as a function of radius $r$ for the dark matter (red dashed lines) and for the total matter (solid black lines). The dashed horizontal line $\gamma=1$ marks the asymptotic value that the NFW profile should reach as $r\to 0$. Results are given for the main BCG progenitor at $z=3$ in the left panels and the final BCG at $z=0$ in the right panels. The upper panels are for run A (contracted) and the lower panels for run B (uncontracted). The BCG progenitors at $z=3$ have inner slopes of $\gamma\sim1.3$ and $\gamma\sim1$ at $5\, \rm{kpc}$ for the dark matter in the contracted and uncontracted models, respectively. In the final systems at $z=0$ these latter inner slopes reach $\gamma\sim0.9$ and $\gamma\sim0.8$. These are depressions of $0.2<\Delta\gamma<0.4$ with respect to the initial slope at this radius. However, as already noted these initial conditions are inconsistent with observations of $z=3$ galaxies assuming a $\Lambda$CDM universe. The behaviour here merely serves to prove that collisionless evolution can reduce the slope of the inner dark matter profile. We still need to test whether this holds for galaxies with more realistic initial configurations, so we turn to results from our weighting scheme.

\begin{figure}
\includegraphics[width=0.5\textwidth,trim=0mm 0mm 0mm 0mm,clip]{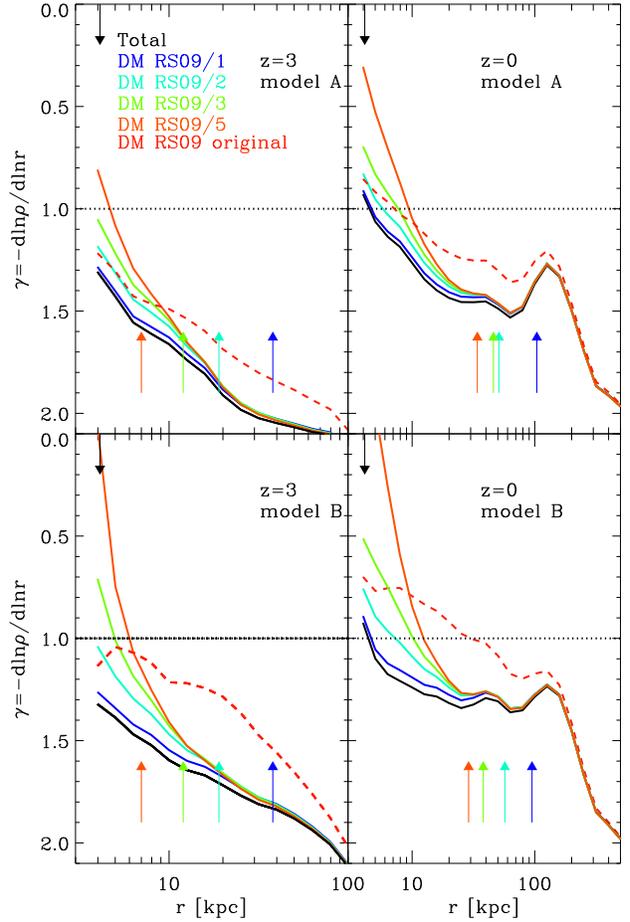}
\caption{Slopes of the total matter and dark matter profiles at $z=3$ and $z=0$ for the four interpretations and for the original RS09 simulations. The top and bottom panels show results for models A and B respectively. The dashed red line is the slope of the dark matter profile in the original RS09 simulations (prior to rescaling). The horizontal dashed black line $\gamma=1$ marks the asymptotic limit for a NFW cusp. The arrows mark the resolution limit $3 \epsilon=4.1 \, \rm{kpc}$. The lower coloured arrows mark the half-light radii for the different representations.}
\end{figure}

\subsection{Dark matter slope evolution for other BCG stellar mass profiles}

Figure 4 shows the initial and final slopes of the dark matter density profiles for each of our reinterpretations of the two simulations using the weighting scheme presented above. The total mass does not of course change from one interpretation to another, only the partition between stellar and dark matter distributions does. Because we have decreased the stellar mass by a factor of ten, the dark matter mass increases to compensate. The distribution of the dark matter varies  however, according to the concentration we assume for the stellar mass. In the RS09/1 interpretation, there is additional dark matter at all radii and $\gamma$ becomes more negative for the dark matter everywhere. In the RS09/5 interpretation however, stars are much more concentrated which results in a shallower initial slope near the centre. The slopes of the dark matter profiles retrieved from the weighting scheme are represented by the solid coloured lines.

The initial dark matter profiles in each interpretation differ from each other in the inner regions where stars contribute significantly to the mass budget ($5-10 \, \rm{kpc}$) and asymptote to the same profiles at larger radii. At $z=3$, the slopes in model A range from $\gamma\sim 1.4$ to $ 1.0$ at 5 kpc, and $\gamma\sim1.6$ to $1.5$ at 10 $\rm{kpc}$ between RS09/1 and RS09/5. At $z=0$ the final slopes take values $\gamma\sim 0.5 $ to $ 1.0$ at $5\, \rm{kpc}$ and $\gamma\sim1.2 $ to $\, 1.0$ at $10 \, \rm{kpc}$. These are changes with respect to the initial slopes of $0.4<\Delta\gamma<0.5$ at both $5 \, \rm{kpc}$ and $10 \, \rm{kpc}$. Note that in a few cases (RS09/4 and RS09/5 model A) the final slopes are already shallower than $\gamma=1$ below $10 \, \rm{kpc}$, a region of interest for observations using stellar kinematics.

For our model B, at $z=3$ the slopes range from $\gamma\sim1.3$ to$\,1.0$ to $\gamma\sim1.6$ to $1.4$ at 5 and 10 kpc respectively. At $z=0$ we also observe significant changes in the inner-slope between $0.3<\Delta\gamma<0.4$ and $0.4<\Delta\gamma<0.6$ at $5$ and $10 \, \rm{kpc}$. We caution that interpretation RS09/5 has a steep slope profile and that for this particular interpretation the slope change is much more significant $\Delta\gamma\sim 0.7$.
 
Thus, we see that the various mergers between galaxies not only change the slopes of the total matter density profiles but also those of the dark matter profiles. We show that even when baryons account for just one percent of the total mass of a galaxy, in agreement with recent abundance matching results, their effect is still significant enough to have an impact on the scales probed by observations, making initial cusps shallower. In the new interpretations, the changes in the slope are more localised to the inner regions ($5-10 \, \rm{kpc}$) of the galaxy when compared to results from the original RS09 simulations where the slope was shallower than $\gamma=1$ out to $\sim30 \, \rm{kpc}$ for model B.

Aspects of this evolution are present in the explanation proposed by \cite{El-Zant2004} for the shallow slopes measured by \cite{Sand2004}: dynamical friction as the galaxies orbit in the cluster heats up the central dark matter and makes cusps shallower. A limitation of this work was the use of massive particles to represent galaxies which can not capture other processes such as mixing and stripping of stars and dark matter. Their shallow cusp was produced by evacuating the central dark matter and replacing it with massive point-like galaxies. The relative distribution of stars and dark matter was thus not followed realistically.

In our numerical experiment, the galaxies are represented by many particles giving them a gravitationally self-consistent phase-space distribution. The numbers of mergers and of galaxies falling to the centre of the cluster are given by the simulation. Using the fact that the galaxies are set up initially to be spherically symmetric and in a steady state, it follows from Jeans' theorem and Eddington's formula that we can construct different equilibrium stellar density distributions out of the total mass distribution of particles. Our simulations show that differences in the weakening of cusps from one interpretation to another are due to differences in the mixing between stars and dark matter. Within the $\Lambda$CDM context, the many collisionless mergers experienced by the BCG produce a weakening of dark matter cusps.

This is a qualitative result and certain aspects of the matter distributions in our simulations remain unrealistic. As a result we do not attempt to reproduce observations of Abell 383 (e.g. the velocity dispersion and luminosity profiles in N11). Even with a fully realistic simulation such a goal would likely be unattainable, given that we have only one realisation of a cluster at our disposition. BCG formation and evolution is tightly coupled to that of the cluster as a whole, so BCGs have a variety of profiles.

In reality the situation is more complicated, galaxies do not form only through collisionless mergers. Disspational processes certainly play an important role in shaping galaxies at earlier times. Baryons dominate the visible regions of the progenitor galaxies and can alter the initial distribution of dark matter in numerous ways: through gas expulsion \citep{Navarro1996}, supernova driven impulsive heating \citep{Pontzen2011}, AGN feedback \citep{Duffy2010} or adiabatic contraction \citep{Blumenthal1986, Gnedin2004, Gnedin2011}.

\section{Conclusions}

We have studied the formation of a $10^{15} \, {\rm{M_{\odot}}}$ cluster in a $\Lambda$CDM Universe, following the evolution of galaxies through collisionless mergers from $z=3$ down to $z=0$. We showed that as a result of mixing between stars and dark matter in dissipationless mergers, initial cusps ($0.8<\gamma<1.3$) can be substantially weakened ($0.3<\gamma<0.9$) at the inner-most resolved radii. Our results indicate that observations of shallow dark matter cusps at the centre of clusters are not necessarily inconsistent with CDM. We find changes of dark matter profile slope at a fixed radius of the order $0.3<\Delta\gamma<0.5$.

Another interesting result from this study is that the evolution in size is stronger than in mass for more compact stellar distributions. We find that the trend moves from $r_{e}\propto M$ for extended galaxies to $r_{e}\propto M^{2}$ for more compact ones. The latter evolutionary trend has in fact been found in a recent observational study by \cite{Dokkum2010}. Our numerical experiments have some serious limitations that we hope to improve on in future work. It would be interesting to use initial conditions based on the observed $z=2$ mass-size relation for galaxies combined with stellar to dark matter ratios consistent with abundance matching arguments in order to study the build-up of BCGs and the fate of the most compact ellipticals in $\Lambda$CDM more reliably. This would allow us to check semi-analytic descriptions of the process in the $\Lambda$CDM context such as that of \cite{Shankar2011}. We are currently investigating this using simulations of the {\sc Phoenix} project \citep{Gao2012}.

\section*{Acknowledgments}
This work was supported by the Marie Curie Initial Training Network
CosmoComp (PITN-GA-2009-238356). MR acknowledges NSF grant NSF 1008454.
\bibliographystyle{mn2e}
\bibliography{master2.bib}{}
\label{lastpage}
\end{document}